\documentclass[useAMS,usenatbib,longnamesfirst,usegraphicx]{mn2e}
\usepackage{txfonts}
\usepackage{color}

\newcommand{\etal}{et al.}
\newcommand{\Chandra}{{\it Chandra}}
\newcommand{\casa}{\textrm{Cas A NS}}

\newcommand{\Ts}{T_{\rm s}}

\begin{document}
\title[Cooling Cas~A Neutron Star]{
Cooling neutron star in the Cassiopeia~A supernova remnant:
Evidence for superfluidity in the core}

\author[P. S. Shternin \etal]{
Peter S. Shternin$^{1,2}$\thanks{E-mail: pshternin@gmail.com},
Dmitry G. Yakovlev$^{1}$,
Craig O. Heinke$^{3}$,\newauthor
Wynn C. G. Ho$^{4}$\thanks{E-mail: wynnho@slac.stanford.edu},
%
%
Daniel J. Patnaude$^{5}$
\\
$^{1}$Ioffe Physical Technical Institute,
Politekhnicheskaya 26, 194021 St.\ Petersburg, Russia \\
$^{2}$St.\ Petersburg State Polytechnical University, Politechnicheskaya 29, 195251, St.\ Petersburg, Russia\\
$^{3}$Department of Physics, University of Alberta, Room 238 CEB,
11322-89 Avenue, Edmonton, AB, T6G 2G7, Canada\\
$^{4}$School of Mathematics, University of Southampton,
Southampton, SO17 1BJ, United Kingdom \\
$^{5}$Smithsonian Astrophysical Observatory, Cambridge, MA 02138,
USA}

\date{Accepted . Received ; in original form}
\pagerange{\pageref{firstpage}--\pageref{lastpage}} \pubyear{2010}
\maketitle \label{firstpage}

\begin{abstract}
According to recent results of \citet{hoheinke09} and
\citet{heinkeho10}, the Cassiopeia~A supernova remnant contains a
young ($\approx 330$ yr old) neutron star (NS) which has carbon
atmosphere and shows noticeable decline of the effective surface
temperature. We report a new (November 2010) \Chandra\ observation
which confirms the previously reported decline rate. The decline is
naturally explained if neutrons have recently become superfluid (in
triplet-state) in the NS core, producing a splash of neutrino
emission due to Cooper pair formation (CPF) process that currently
accelerates the cooling. This scenario puts stringent constraints on
poorly known properties of NS cores: on density dependence of the
temperature $T_\mathrm{cn}(\rho)$ for the onset of neutron
superfluidity [$T_\mathrm{cn}(\rho)$ should have a wide peak with
maximum $\approx (7-9)\times 10^8$~K], on the reduction factor $q$
of CPF process by collective effects in superfluid matter ($q >
0.4$), and on the intensity of neutrino emission before the onset of
neutron superfluidity (30--100 times weaker than the standard
modified Urca process). This is serious evidence for nucleon
superfluidity in NS cores that comes from observations of cooling
NSs.
\end{abstract}


\section{Introduction} \label{sec:intro}

It is well known that NS cores contain superdense matter whose
properties are still uncertain (see, e.g.,
\citealt{hpy07,lattimerprakash07}). One can explore these properties
by studying the cooling of isolated NSs \citep[see, e.g.,][for
review]{pethick92,yakovlevpethick04,pageetal06,pageetal09}.

We analyse observations of the NS in the supernova remnant
Cassiopeia~A (Cas~A). The distance to the remnant is
$d=3.4^{+0.3}_{-0.1}$~kpc \citep{reedetal95}. The Cas~A age is
reliably estimated as $t \approx 330\pm 20$~yr from observations of
the remnant expansion \citep{fesenetal06}. The compact central
source was identified in first-light \Chandra\ X-ray observations
\citep{tananbaum99} and studied by
\citet{pavlovetal00,chakrabartyetal01,pavlovluna09} but its nature
has been uncertain. The fits of the observed X-ray spectrum with
magnetized or non-magnetized hydrogen atmosphere models or with
black-body spectrum revealed too small size of the emission region
(could be hot spots on NS surface although no pulsations have been
observed, e.g., \citealt{pavlovluna09}).

Recently \citet{hoheinke09} have shown that the observed spectrum is
successfully fitted taking a carbon atmosphere model with a low
magnetic field ($B \lesssim 10^{11}$ G). The gravitational mass of
the object, as inferred from the fits, is $M \approx
1.3-2\,M_\odot$, circumferential radius $R \approx 8-15$ km, and the
non-redshifted effective surface temperature $\Ts \sim 2 \times
10^6$~K \citep{yakovlevetal10}. These parameters indicate that the
compact source is an NS with the carbon atmosphere. It emits thermal
radiation from the entire surface and has the surface temperature
typical for an isolated NS. It is the youngest in the family of
observed cooling NSs.

\citet{yakovlevetal10} compared these observations with the NS
cooling theory. The authors concluded that the \casa\ has already
reached the stage of internal thermal relaxation. It cools via
neutrino emission from the stellar core; its neutrino luminosity is
not very different from that provided by the modified Urca process.

Following \citet{hoheinke09}, \citet{heinkeho10} analysed \Chandra\
observations of the \casa\ during 10 years and found a steady
decline of $\Ts$ by about 4\%. They interpret it as direct
observation of \casa\ cooling, the phenomenon which has never been
observed before for any isolated NS. These results are confirmed by
new observations we report below. We interpret them as a
manifestation of neutron superfluidity in the \casa.

When this paper was nearly completed we became aware of the paper by
\citet{pageNew10} who proposed similar explanation of the \casa\
observations. However the two papers are different in details, and
can be regarded as complementary. In particular, we discuss the
dependence of cooling curves on the poorly known efficiency of
neutrino emission due to CPF process and the possibility to
interpret observations of all cooling stars by one model of
superdense matter. It is important that we report the new
observation.

\section{New Chandra Observations}\label{S:obs}

We use the \Chandra\ data on the \casa, discussed and fitted by
\citet{heinkeho10}, and add one new data point.  Briefly, the
analysed data include the \Chandra\ ACIS-S observations of Cas A
longer than 5 ks.  In order to ensure that all considered data are
directly comparable, we
take
only
the `directly comparable data' of \citet{heinkeho10}.
We exclude the subarray observation of Cas A \citep{pavlovluna09},
since the pileup properties \citep{Davis01} of this spectrum differ
from the others, and those observations in which the Cas A NS
dithered over bad pixels (most of the 2004 observations).

The new data point is produced from two ACIS-S observations of Cas A
(\citealt{Patnaude10}: ObsIDs 10936, 13177) taken on 2010 October~31
and 2010 November~2 for 33 and 17 ks (respectively) and telemetered
in GRADED mode (as done for previous Cas A observations).   We used
{\small CIAO 4.2} and {\small CALDB 4.2.1} to reprocess the data and
produce response functions, correcting for the time-dependent ACIS
quantum efficiency degradation and gain changes, but not for the
charge-transfer inefficiency (as this cannot be modeled with GRADED
data).  We extracted source spectra with a 4 pixel (2.37") radius
region, and background spectra from an annulus of 5 to 8 pixels, as
in \citet{heinkeho10}.  We combined the two new observations into
one set of spectra and responses (with an effective date of 2010
November 1), and grouped the spectrum by 200 counts/bin.

We fitted these spectra simultaneously, forcing the NS mass and
radius, along with the distance and $N_\mathrm{H}$, to be the same
as in the best fit of \citet{heinkeho10} and \citet{yakovlevetal10},
and finding the 1$\sigma$ errors on the surface temperature at each
epoch. This fitting is designed to cleanly define the relative
variation in the temperature, separating this question from the
absolute uncertainty in the temperature (described in detail in
\citealt{yakovlevetal10}). The fitted values of the non-redshifted
effective surface temperature $T_\mathrm{s}$ (Table 1) are slightly
different from those reported by \citet{heinkeho10} for the
2004--2009 observations, but within the 1$\sigma$ errors.  The key
result is that the new observation confirms and extends the cooling
trend seen in \citet{heinkeho10}.

\begin{table}
\caption[]{Carbon atmosphere spectral fits, using the best spectral
fit ($M$, $R$, $N_\mathrm{H}$) of \citet{heinkeho10} and
\citet{yakovlevetal10}, with the addition of 2010 data.  Epoch dates
are for the midpoints of the observations, or weighted midpoints of
merged datasets. Temperature errors are $1\sigma$ confidence for a
single parameter.}
\begin{tabular}{lccc}
{\textbf{Epoch}} & {\textbf{Exposure}}  & {$\log T_\mathrm{s}$} &
{\textbf{ObsID(s)}} \\
 (Year) &  ks & K  & \\
 \hline\hline
2000.08  &  50.56  & 6.3258$^{+0.0019}_{-0.0019}$ &  114 \\
2002.10  &  50.3    & 6.3237$^{+0.0018}_{-0.0018}$ &  1952 \\
2004.11  &  50.16  & 6.3156$^{+0.0019}_{-0.0019}$ &  5196 \\
2007.93  &  50.35  & 6.3108$^{+0.0019}_{-0.0019}$ &  9117, 9773 \\
2009.84  &  46.26  & 6.3087$^{+0.0018}_{-0.0018}$ &  10935, 12020 \\
2010.83  &  49.49  &  6.3060$^{+0.0019}_{-0.0018}$ & 10936, 13177
\end{tabular}
\end{table}

\section{Second temperature drop}
\label{s:relax}
%

\begin{figure*}
\includegraphics[width=0.7\textwidth]{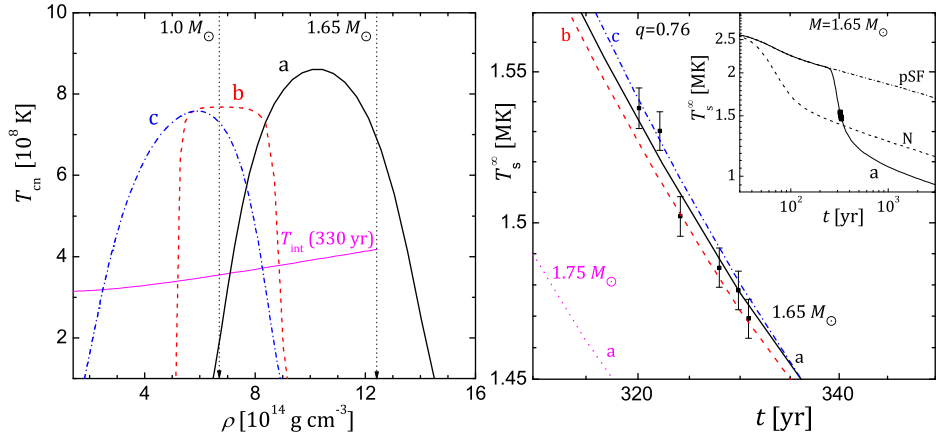}
\caption{(Color on line) {\it Left:} Models (a), (b) and (c) for
critical temperature of triplet-state neutron pairing in NS core.
Vertical dotted lines show central densities of NSs with $M=1$ and
1.65~$M_\odot$ (for APR EOS). Thin solid line is the temperature
profile in the 1.65~$M_\odot$ star of age 330 yr with neutron
superfluidity (a). {\it Right:} Lines (a), (b) or (c) show cooling
of the 1.65~$M_\odot$ star with strong proton superfluidity and
moderate neutron superfluidity (a), (b) or (c) in the core
($q=0.76$) compared with observations of \casa\ cooling. The dotted
line is the same for the 1.75~$M_\odot$ star and neutron
superfluidity (a). The line (a) for $M=1.65\,M_\odot$ is also shown
in the inset over longer time span. Lines pSF and N in the inset are
calculated for $M=1.65\,M_\odot$ neglecting, respectively, neutron
superfluidity and entire nucleon superfluidity in the core.
\label{fig:Tc} }
\end{figure*}

The observed surface temperature decline is too steep and cannot be
described by the theory discussed by \citet{yakovlevetal10} (who did
not analyse the decline itself). \citet{heinkeho10} suggested that
\casa\ undergoes the last years of the internal crust-core
relaxation accompanied by a pronounced surface temperature drop.
However, the theory predicts (e.g.,
\citealt{lattimeretal94,gnedinetal01,yakovlevetal10}) a shorter
relaxation, lasting typically $\lesssim 100$ yr. We propose another
interpretation based on the effects of superfluidity in NS cores, in
line with the work of \citet{gusakovetal04}.

Neutrons, protons (and other baryons if present) in NS interiors can
be superfluid (due to Cooper pairing). Free neutrons in the inner
crust and protons in the core undergo Cooper pairing in spin-singlet
state, while neutrons in the core can pair in spin-triplet state.
Critical temperatures of superfluidity onset $T_\mathrm{c}(\rho)$
are very model dependent (as reviewed, e.g., by
\citealt{pageetal04}).

Our idea is that the initial crust-core relaxation in \casa\ is
over, leaving the star sufficiently warm. We assume further that not
too strong triplet-state neutron superfluidity appeared in the core
some time ago. It initiates a splash of neutrino emission due to CPF
process producing the second $\Ts$ drop that mimics the second
(delayed) thermal relaxation. The second drop is mainly regulated by
(i) the density dependence $T_\mathrm{cn}=T_\mathrm{cn}(\rho)$ of
the critical temperature for the
neutron pairing in the NS core, (ii) the reduction factor $q$ of CPF
process by collective effects in superfluid matter and (iii) the
neutrino luminosity prior to the onset of neutron superfluidity.
These properties are very model dependent and poorly known but can
be constrained from the \casa\ observations.

The CPF process was first predicted by \citet{frs76} for
singlet-state pairing neglecting collective superfluid effects.
Collective effects can suppress the process ($q<1$) as was first
noticed by \citet{leinson01} and calculated later by
\citet{lp06,sedrakianetal07,kv08,sr09,leinson10}; some discussion is
also given by \citet{pageetal09};
the results are controversial. The main attention has been paid to
CPF process due to singlet-state pairing of neutrons which has been
found to be strongly suppressed ($q \ll 1$). We are interested in
triplet-state pairing in which case the suppression is thought to be
less dramatic.
If the latter CPF process were not affected by collective effects,
then 24\% of neutrino emissivity would go through the vector weak
interaction channel, while the rest (76\%) would go through the
axial-vector channel. Collective effects suppress the vector channel
almost completely but the axial-vector channel survives. Exact value
of $q$ is debatable (the lowest estimate $q \approx 0.19$ is given
by \citealt{leinson10}). Instead of relying on any specific model,
we consider $q$ as a free parameter.

We illustrate this idea by cooling simulations of NSs with nucleon
cores. We take the Akmal-Pandharipande-Ravenhall (APR) equation of
state (EOS) in the core \citep{akmaletal98}. By APR we mean the
parametrization of APR results by \citet{hhj99} -- the version APR~I
proposed by \citet{gusakovetal05}. The maximum mass of a stable NS
for this EOS is $M_\mathrm{max}=1.929 M_\odot$; the powerful direct
Urca process of neutrino emission is open in stars with
$M>M_\mathrm{DU}=1.829 M_\odot$. Let the direct Urca process and
even less efficient modified Urca process be either not allowed or
strongly suppressed in the \casa. Otherwise the star would be too
cold after the initial crust-core relaxation; we would be unable to
significantly speed up its cooling by the CPF process. To suppress
Urca processes we assume the presence of strong proton superfluidity
in the core, with critical temperature $T_\mathrm{cp}(\rho) \gtrsim
(2-3) \times 10^9$~K; exact values of $T_\mathrm{cp}$ are
unimportant. It occurs within a few days after the star formation.
The proton CPF neutrino emission does not influence the cooling
even if this emission were not affected by
collective effects \citep[e.g.,][]{yakovlevetal01}. As long as
neutrons are non-superfluid, the neutrino emission is mainly
generated in neutron-neutron bremsstrahlung process. It is weak and
leaves the \casa\ sufficiently warm before the second temperature
drop. Unless the contrary is indicated, we consider NSs with
ordinary (non-accretted, non-magnetized) heat blanketing envelopes
(e.g., \citealt{potekhinetal97,potekhinetal03}) and neglect
superfluid effects in the stellar crust (which weakly affect the
cooling after the crust-core relaxation; e.g.,
\citealt{yakovlevetal01}). Note, that a splash of CPF neutrinos
makes the star slightly non-isothermal, with the cooling slightly
dependent on the thermal conductivity in the core, but the
isothermal state is soon restored.

\begin{figure*}
\includegraphics[width=0.7\textwidth]{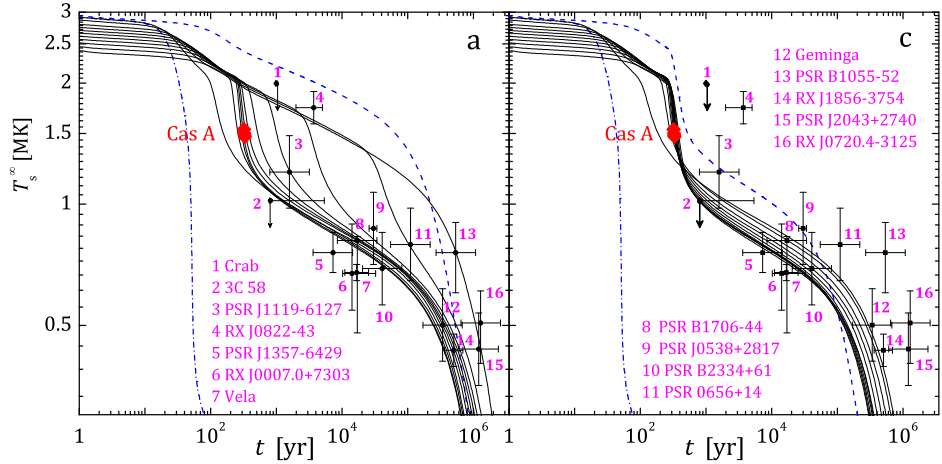}
\caption{(Color on line)
Sequences of (solid) cooling curves for NSs of masses from $1\,
M_\odot$ to $M_\mathrm{max}$ (through $0.1 M_\odot$) with strong
proton superfluidity and moderate neutron superfluidity (a) ({\it
left}) or (c) ({\it right}) in the core ($q=0.76$) compared with
observations of isolated NSs. Dashed lines refer to warmest stars of
these types -- $1\,M_\odot$ stars with the carbon surface layer of
mass $10^{-8}\,M_\odot$. Dash-and-dotted lines refer to coolest
$M_\mathrm{max}$ stars without proton superfluidity in the inner
core.
\label{fig:Cool} }
\end{figure*}

\begin{figure}
\includegraphics[width=0.8\columnwidth]{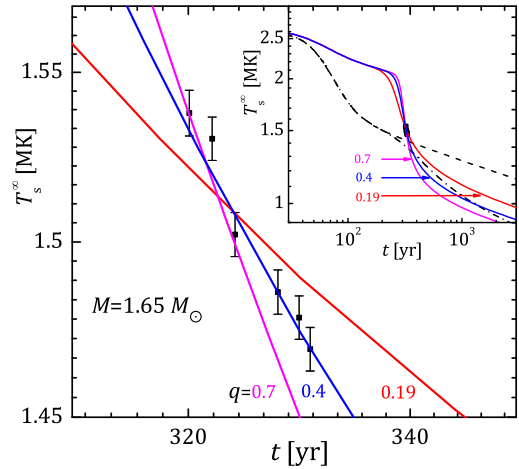}
\caption{(Color on line)
Same as on the right panel of Fig.~\ref{fig:Tc} but for constant
$T_\mathrm{cn}$ over the core at three values $q=0.19$
($T_\mathrm{cn}=7.55 \times 10^8$ K), 0.4 ($7.2 \times 10^8$~K) and
0.7 ($7 \times 10^8$~K). The inset shows the same cooling curves but
over larger range of ages, together with the dashed curve for
non-superfluid star and the dash-and-dot curve for the star without
proton superfluidity but with neutron superfluidity at
$T_\mathrm{cn}=4.3 \times 10^8$~K.
\label{fig:Reduce} }
\end{figure}

Illustrative results are shown in Figs.\ \ref{fig:Tc},
\ref{fig:Cool} and \ref{fig:Reduce}. The left panel of Fig.\
\ref{fig:Tc} gives three models of $T_\mathrm{cn}(\rho)$. We do
not rely on any specific theoretical model but consider three
phenomenological curves (a), (b) and (c) (locating neutron
superfluidity at progressively lower densities). In our case the
maximum of $T_\mathrm{cn}(\rho)$ is strictly constrained to
$T_{\rm cn\; max}\approx (7-9) \times 10^8$~K. Higher or lower
$T_\mathrm{cn}(\rho)$ peaks would start the second temperature
drop in the \casa\ earlier or later than required by the
observations (or even completely wash out this drop in a very
young or old star). The $T_\mathrm{cn}(\rho)$ profiles should not
be too narrow and the reduction factor $q$ should not be too small
(otherwise the second temperature drop would be weak). Other plots
give theoretical cooling curves $\Ts^\infty(t)$ (i.e., the
redshifted surface temperature versus $t$).

The right panel of Fig.\ \ref{fig:Tc} gives three cooling curves for
the $1.65\,M_\odot$ star of nearly \casa\ age with
$T_\mathrm{cn}(\rho)$ models (a)--(c) from the left panel and with
$q=0.76$ (axial vector channel is not suppressed). The curves are
compared with the \casa\ data. All three superfluidity models agree
with these data. Taking superfluidity (a) and increasing the stellar
mass to $1.75\,M_\odot$ makes the CPF process more important and
starts the second temperature drop earlier (the dotted curve), in
disagreement with the data. However, we could easily readjust
(slightly decrease) $T_\mathrm{cn}(\rho)$ and explain the \casa\
data with the $1.75\,M_\odot$ model. Therefore, we can successfully
fit the data for a range of NS masses and take the $1.65\,M_\odot$
star as an example. The inset shows the cooling curve (a)
($M=1.65\,M_\odot$) over longer time scale. The second temperature
drop at $t\sim 250$~yr is clearly pronounced here. Also, we show the
cooling of non-superfluid star (curve N) and the star with proton
superfluidity alone (curve pSF). Without the CPF process, the slopes
of both curves are much smaller than required.

In Fig.\ \ref{fig:Cool} we test two neutron superfluidity models,
(a) and (c), against observations of other isolated NSs.
Observational data are taken from references cited in
\citet{yakovlevetal08} and \citet{kaminkeretal09} with the exception
of PSR J0007+7303. The data on the latter source are taken from
\citet{caraveoetal10} and \citet{linetal10}. Note that in similar
fig.\ 5 of \citet{yakovlevetal10} source labels 12, 13 and 14 should
refer to RX J1856--3754, Geminga and PSR B1055--52, respectively.

We make a reasonable assumption that all NSs have different masses
but the same physics of matter in their cores. We present cooling
curves (solid lines) for a number of masses (from top to bottom),
from $1\, M_\odot$ to $M_\mathrm{max}$. Recall, that we use the
model of strong proton superfluidity in the entire core that
switches off Urca processes in all stars. With this assumption,
neutron superfluidity (a) allows us to explain almost all sources,
except for warmest and coolest ones (for their ages). However,
neutron superfluidity (c) produces too strong CPF neutrino emission
in low-mass stars and cannot explain the majority of warmer stars,
although it agrees with the \casa\ data.

Furthermore, we can rise the cooling curves of low-mass, not too old
NSs assuming they have more heat transparent heat-blanketing
envelopes of light elements . For example, the dashed curves in
Fig.\ \ref{fig:Cool} are calculated for $M=1\,M_\odot$ NS with the
carbon envelope of mass $10^{-8}\,M_\odot$. We can also lower the
cooling curves of massive ($M_\mathrm{max}$) stars: the dash-dotted
lines are computed by taking more realistic models for proton
superfluidity, with $T_\mathrm{cp}(\rho)$ going down at high
densities. This opens direct Urca process in the inner core and
gives coldest possible NSs. Therefore, we can really explain all the
data with superfluidity (a), but not with superfluidity~(c).

The second temperature drop is known in the NS cooling theory (e.g.,
\citealt{kaminkeretal01}). Cooling models like (a) and (c) in Fig.\
\ref{fig:Cool} have been analysed by \citet{gusakovetal04} with the
same conclusion that models like (a) can explain observations of all
isolated NSs. Similar models of NSs with nucleon cores, where direct
Urca is forbidden but CPF operates, were used as the basis of the
minimal cooling theory (\citealt{pageetal04,pageetal06,pageetal09})
although that theory employs selected $T_\mathrm{c}(\rho)$ profiles,
most favorable by the theory of nucleon superfluidity. Now we see
that the model of \citet{gusakovetal04} is also suitable to explain
the \casa\ data.

Finally, Fig.\ \ref{fig:Reduce} demonstrates the effect of
suppression of CPF neutrino emission in the axial-vector channel. To
maximize the CPF emission we take constant $T_\mathrm{cn}$ over the
core. It produces especially strong splash of CPF neutrinos when the
second temperature drop starts. Three solid lines are the cooling
curves for the $1.65\,M_\odot$ star calculated at $q$=0.7, 0.4 and
0.19. Otherwise the conditions are the same as in the right panel of
Fig.\ \ref{fig:Tc}. With our constant $T_\mathrm{cn}$, CPF neutrino
emission at $q=0.7$ is too strong; it gives faster \casa\ cooling
than required by observations. The case $q=0.4$ now agrees with the
observations. Smaller $q=0.19$ gives slower cooling that cannot
explain the data. Notice that the $T_\mathrm{cn}$=const model is
hardly realistic. For more realistic $T_\mathrm{cn}(\rho)$ profiles
we can reconcile theory with the \casa\ data at $q > 0.4$. This
gives a useful restriction on the uncertain theoretical parameter
$q$.

In the inset of Fig.\ \ref{fig:Reduce} we show the same three
cooling curves over larger range of ages. In addition, we plot the
same dashed line for non-superfluid star as in the right panel of
Fig.\ \ref{fig:Tc}, and another dot-and-dashed line for the star
with neutron superfluidity alone with $T_\mathrm{cn}=4.3 \times
10^8$~K  and $q=0.76$. The latter superfluidity triggers a splash
of CPF neutrinos, but the main modified Urca neutrino emission is
too strong and the splash cannot produce a steep $\Ts$ decline
required by the observations. Adding a carbon surface layer of
mass $\sim 10^{-12}\,M_\odot$, we could raise the latter curve to
the \casa\ level but would be unable to reproduce the cooling
slope. Our calculations show that the modified Urca emission
should be suppressed at least by a factor of 30 (for the most
efficient CPF emission with $q=0.76$ and constant $T_\mathrm{cn}$)
to get the required slope. Taking smaller $q$ or narrower
$T_\mathrm{cn}(\rho)$-profile would require stronger suppression
of the modified Urca process.

\section{Conclusions} \label{sec:concl}

We report a new (November, 2010) \Chandra\ observation of the young
\casa\ that confirms the observed \citep{heinkeho10} steady decline
of the surface temperature $\Ts$ (by 4\% over 10 years). We propose
a natural explanation of the observed decline. We assume that the
\casa\ underwent the traditional internal crust-core relaxation some
time ago and now demonstrates the second temperature drop due to the
onset of triplet-state neutron superfluidity in its core and
associated neutrino emission. We can explain the \casa\ observations
under the following conditions:

\begin{itemize}

\item
The maximum critical temperature for triplet-state pairing of
neutrons should be $T_{\rm cn\; max} \approx (7-9) \times 10^8$~K.
Otherwise the second temperature drop occurs earlier or later than
required by observations.

\item
The $T_\mathrm{cn}(\rho)$ profile over the NS core should be rather
wide for the CPF neutrino emission to gain enough strength.

\item
For the same reason the suppression of the CPF process by collective
effects cannot be too strong ($q > 0.4$).

\item
The neutrino emission of the star before the second temperature
drop should be 30--100 times lower than due to the modified Urca
process (e.g., the modified Urca can be suppressed by strong
proton superfluidity). Otherwise the second temperature drop would
not be pronounced.

\end{itemize}

When these criteria are met, we can still locate
$T_\mathrm{cn}(\rho)$ profiles in different parts of the NS core.
If, however, we wish to explain all current observations of isolated
NSs with one and the same $T_\mathrm{cn}(\rho)$-profile, we will be
forced to push this profile deeper in the core (Fig.\
\ref{fig:Cool}). Alternatively, we could employ broader profiles but
with density-dependent factor $q$ (which can increase within the
core as for singlet-state pairing, e.g., \citealt{kv08}). This would
shift the efficiency of the CPF process to higher $\rho$ in
superfluid matter.

We have taken one EOS and focused on $1.65 \, M_\odot$ neutron
star model but our basic conclusions will not change for a large
variety of EOSs and masses $M$. 
For instance, taking the same EOS we have considered the \casa\
models with $M$ from $1.4\,M_\odot$ to $1.9\,M_\odot$. By slightly
changing $T_\mathrm{cn}(\rho)$ profiles we are able to explain the
data for any $M$ from this range.

Our calculations indicate that the second temperature drop lasts for
a few tens of years and \casa\ is at its active CPF neutrino
emission stage. These models would be inconsistent with a sharp stop
of the temperature decline in a few years, which can be verified
with future observations.

After the second temperature drop the \casa\ is expected to become a
rather cold slowly cooling NS.


\section*{acknowledgements}

We are grateful to A.~Y.~Potekhin for critical remarks. PSS and DGY
acknowledge support from Rosnauka (grant NSh 3769.2010.2) and
Ministry of Education and Science of Russian Federation (contract
11.G34.31.0001 with SPbSPU and Leading Scientist G.~G.~Pavlov). WCGH
acknowledges support from the Science and Technology Facilities
Council (STFC) in the United Kingdom through grant number
PP/E001025/1. PSS acknowledges support of the Dynasty Foundation and
RF Presidential Program MK-5857.2010.2. COH acknowledges support
from the Natural Sciences and Engineering Research Council (NSERC)
of Canada.


\label{lastpage}

\end{document}